\documentclass[11pt,english]{article}
\usepackage[T1]{fontenc}
\usepackage{geometry}
\usepackage{amsmath}
\usepackage{setspace}
\usepackage{amssymb}
\usepackage{esint}

\makeatletter
\newcommand{\lyxaddress}[1]{
\par {\raggedright #1
\vspace{1.4em}
\noindent\par}
}

\usepackage{babel}
\makeatother

\begin{document}

\title{Magnetohydrodynamics in Presence of Electric and Magnetic charges}

\author{P. S. Bisht$^{\text{(1,2)}}$, Pushpa$^{\text{(1)}}$ and O. P. S.
Negi$^{\text{(1)}}$}

\maketitle

\lyxaddress{\begin{center}
$^{\text{(1)}}$Department of Physics\\
 Kumaun University\\
S. S. J. Campus \\
Almora-263601 (Uttarakhand) India
\par\end{center}}

\lyxaddress{\begin{center}
$^{\text{(2)}}$ Institute of Theoretical Physics\\
 Chinese Academy of Sciences\\
P. O. Box 2735\\
 Beijing 100080, P. R. China\\

\par\end{center}}

\lyxaddress{\begin{center}
Email- ps\_bisht 123@rediffmail.com \\
pushpakalauni60@yahoo.co.in\\
ops\_negi@yahoo.co.in
\par\end{center}}

\begin{abstract}
Starting with the generalized electromagnetic field equations of dyons,
we have discussed the theory of magnetohydrodynamics (MHD) of plasma
for particles carrying simultaneously the electric and magnetic charges
(namely dyons). It is shown that the resultant system supports the
electromagnetic duality of dyons. Consequently the frequency of dyonic
plasma has been obtained and it is emphasized that there is a different
plasma frequency for each species depending on wave number $k$. For
$k$ to be real, only those generalized electromagnetic waves are
allowed to pass, for which the usual frequency is greater than the
plasma frequency (i.e. $\omega>\omega_{p})$. It is shown that the
plasma frequency sets the lower cuts for the frequencies of electromagnetic
radiation that can pass through a plasma . Accordingly the ohm's law
has been reestablished to derive the plasma oscillation equation as
well as the magetohydrodynamic wave equation and the energy of dyons
in unique and consistent manner.

\textbf{PACS NO}: 14.80.Hv (Magnetic Monopoles), 52.30.Cv (Magnetohydrodynamics
-including electron magnetohydrodynamics)

\end{abstract}

\section{Introduction}

~\ \ \ \ \ \ \ \ \ \  Magneto hydrodynamics (MHD) is a
branch of the science of the dynamics of matter moving in an electromagnetic
field \cite{key-1} and thus provides one of the most useful fluid
models, focusing on the global properties of plasma. In a series of
papers, we \cite{key-2,key-3,key-4,key-5,key-6,key-7} have undertaken
the study of dual electrodynamics and superluminal electromagnetic
fields, developed the quaternionic formulation of dyons in isotropic
homogeneous, chiral and inhomogeneous media and obtained the solutions
for the classical problem of moving dyon in unique and consistent
way. Coceal et. al. \cite{key-8} derived consistently the duality
invariant magnetohydrodynamics and their dyonic solutions. Keeping
in view, in this paper, we have discussed the theory of magnetohydrodynamics
(MHD) of plasma for particles carrying simultaneously the electric
and magnetic charges (namely dyons). It is shown that the resultant
system supports the electromagnetic duality of dyons. Consequently
the frequency of dyonic plasma has been obtained and it is emphasized
that there is a different plasma frequency for each species depending
on wave number $k$. For $k$ to be real, only those generalized electromagnetic
waves are allowed to pass, for which the usual frequency is greater
than the plasma frequency (i.e. $\omega>\omega_{p})$. It is shown
that the plasma frequency sets the lower cuts for the frequencies
of electromagnetic radiation that can pass through a plasma . Accordingly
the ohm's law has been reestablished to derive the plasma oscillation
equation as well as the magetohydrodynamic wave equation and the energy
of dyons in unique and consistent manner.

\section{Fields Associated with dyons}

Postulating the existence of magnetic monopoles, the generalized Dirac
Maxwell's (GDM) equations \cite{key-9} are expressed in SI units
$(c=\hbar=1)$ as 

\begin{eqnarray}
\overrightarrow{\nabla.} & \overrightarrow{E}= & \rho_{e};\nonumber \\
\overrightarrow{\nabla.} & \overrightarrow{H} & =\rho_{m};\nonumber \\
\overrightarrow{\nabla}\times\overrightarrow{E}=-\frac{\partial\overrightarrow{H}}{\partial t} & - & \overrightarrow{j_{m}};\nonumber \\
\overrightarrow{\nabla}\times\overrightarrow{H}=\frac{\partial\overrightarrow{E}}{\partial t} & + & \overrightarrow{j_{e}};\label{eq:1}\end{eqnarray}
where $\rho_{e}$ and $\rho_{m}$ are respectively the electric and
magnetic charge densities, $\overrightarrow{j_{e}}$ and $\overrightarrow{j_{m}}$
are the corresponding current densities , $\vec{E}$ is electric field
and $\vec{H}$ is magnetic field. GDM equations (\ref{eq:1}) are
invariant not only under Lorentz and conformal transformations but
also invariant under the following duality transformations \cite{key-2,key-10,key-11},

\begin{eqnarray}
\overrightarrow{E}=\overrightarrow{E} & \cos\theta+\overrightarrow{H} & \sin\theta;\nonumber \\
\overrightarrow{H}=-\overrightarrow{E} & \sin\theta+\overrightarrow{H} & \cos\theta.\label{eq:2}\end{eqnarray}
For a particular value of $\theta$=$\frac{\Pi}{2}$, equation (\ref{eq:2})
reduces to

\begin{eqnarray}
\overrightarrow{E}\rightarrow\,\overrightarrow{H}, & \overrightarrow{H}\rightarrow- & \overrightarrow{E},\label{eq:3}\end{eqnarray}
which can be written as \begin{eqnarray}
\left(\begin{array}{c}
\overrightarrow{E}\\
\overrightarrow{H}\end{array}\right)\Rightarrow & \left(\begin{array}{cc}
0 & 1\\
-1 & 0\end{array}\right) & \left(\begin{array}{c}
\overrightarrow{E}\\
\overrightarrow{H}\end{array}\right).\label{eq:4}\end{eqnarray}
If we apply the transformation (\ref{eq:3}) and (\ref{eq:4}) along
with the following duality transformations for current i.e;\begin{eqnarray}
\rho_{e}\rightarrow\rho_{m},\rho_{m}\rightarrow-\rho_{e}\Longleftrightarrow & \left(\begin{array}{c}
\rho_{e}\\
\rho_{m}\end{array}\right)\Rightarrow\left(\begin{array}{cc}
0 & 1\\
-1 & 0\end{array}\right) & \left(\begin{array}{c}
\rho_{e}\\
\rho_{m}\end{array}\right).\label{eq:5}\end{eqnarray}
Differential equation (\ref{eq:1}) are the generalized field equations
of dyons and the corresponding electric and magnetic fields are then
called generalized electromagnetic field of dyons are expressed in
the following differential form in terms of two potentials \cite{key-2}:

\begin{eqnarray}
\overrightarrow{E}=-\overrightarrow{\nabla} & \phi_{e}-\frac{\partial\overrightarrow{A}}{\partial t}- & \overrightarrow{\nabla}\times\overrightarrow{B};\label{eq:6}\end{eqnarray}
\begin{eqnarray}
\overrightarrow{H}=-\overrightarrow{\nabla}\phi_{g}-\frac{\partial\overrightarrow{B}}{\partial t} & + & \overrightarrow{\nabla}\times\overrightarrow{A};\label{eq:7}\end{eqnarray}
where $\left\{ A^{\mu}\right\} $=$\left\{ \phi_{e},\,\overrightarrow{A}\right\} $
and $\left\{ B^{\mu}\right\} $=$\left\{ \phi_{g},\,\overrightarrow{B}\right\} $are
the two four - potentials associated with electric and magnetic charges.
Let us define the complex vector field $\overrightarrow{\psi}$ in
the following form,

\begin{eqnarray}
\overrightarrow{\psi}= & \overrightarrow{E}-i & \overrightarrow{H},\label{eq:8}\end{eqnarray}
equations (\ref{eq:6},\ref{eq:7}) and (\ref{eq:8}) , thus give
rise to the following relation between generalized field and the components
of the generalized four - potential as;

\begin{eqnarray}
\overrightarrow{\psi}=-\frac{\partial\overrightarrow{V}}{\partial t}-\overrightarrow{\nabla}\phi-i\overrightarrow{\nabla} & \times & \overrightarrow{V}.\label{eq:9}\end{eqnarray}
Here $\left\{ V_{\mu}\right\} $ is the generalized four - potential
of dyons and defined as ;\begin{eqnarray}
\left\{ V_{\mu}\right\}  & = & \left\{ \phi,-\overrightarrow{V}\right\} ;\label{eq:10}\end{eqnarray}
where

\begin{eqnarray}
\phi=\phi_{e}- & i & \phi_{m};\label{eq:11}\end{eqnarray}
and

\begin{eqnarray}
\overrightarrow{V}= & \overrightarrow{A}-i & \overrightarrow{B}.\label{eq:12}\end{eqnarray}
If we apply the transformation (\ref{eq:4}) and (\ref{eq:5}) the
following duality transformation for potential is obtained i.e.

\begin{eqnarray}
\overrightarrow{A}\rightarrow\overrightarrow{B},\,\overrightarrow{B}\rightarrow-\overrightarrow{A}\Rightarrow\left(\begin{array}{c}
\overrightarrow{A}\\
\overrightarrow{B}\end{array}\right)= & \left(\begin{array}{cc}
0 & 1\\
-1 & 0\end{array}\right) & \left(\begin{array}{c}
\overrightarrow{A}\\
\overrightarrow{B}\end{array}\right);\label{eq:13}\end{eqnarray}

\begin{eqnarray}
\phi_{e}\rightarrow\phi_{m},\,\phi_{e}\rightarrow-\phi_{m}\Rightarrow\left(\begin{array}{c}
\phi_{e}\\
\phi_{m}\end{array}\right)= & \left(\begin{array}{cc}
0 & 1\\
-1 & 0\end{array}\right) & \left(\begin{array}{c}
\phi_{e}\\
\phi_{m}\end{array}\right).\label{eq:14}\end{eqnarray}
Maxwell field equation (\ref{eq:1}) may then be written in terms
of generalized field $\overrightarrow{\psi}$ as;

\begin{eqnarray*}
\overrightarrow{\nabla}. & \overrightarrow{\psi} & =\rho;\end{eqnarray*}

\begin{eqnarray}
\overrightarrow{\nabla} & \times\overrightarrow{\psi}=-i\overrightarrow{J}-i & \frac{\partial\overrightarrow{\psi}}{\partial t};\label{eq:15}\end{eqnarray}
where $\rho$ and $\vec{J}$ are the generalized charge and current
source densities of dyons and given by;

\begin{eqnarray}
\rho=\rho_{e} & -i & \rho_{g};\label{eq:16}\end{eqnarray}

\begin{eqnarray}
\overrightarrow{J}=\overrightarrow{j_{e}} & -i & \overrightarrow{j_{m}}.\label{eq:17}\end{eqnarray}
Here, we may write the tensorial form of generalized Maxwell Dirac
equation of dyons as,

\begin{eqnarray}
F_{\mu\nu,\nu} & = & j_{\mu}^{e};\nonumber \\
F_{\mu\nu,\nu}^{d} & = & j_{\mu}^{m};\label{eq:18}\end{eqnarray}
where\begin{eqnarray*}
\left\{ j_{\mu}^{e}\right\} =\left\{ \rho_{e},\overrightarrow{\, j_{e}}\right\}  & and\,\,\,\left\{ j_{\mu}^{m}\right\} = & \left\{ \rho_{m},\,\overrightarrow{j_{m}}\right\} .\end{eqnarray*}
 Defining the generalized field tensor of dyons as;

\begin{eqnarray}
G_{\mu\nu,\nu} & =F_{\mu\nu}-i & F_{\mu\nu}^{d}\label{eq:19}\end{eqnarray}
one can directly obtained the following generalized field equation
of dyon i.e;

\begin{eqnarray}
G_{\mu\nu,\nu} & = & J_{\mu};\nonumber \\
G_{\mu\nu,\nu} & = & 0;\label{eq:20}\end{eqnarray}
where \begin{eqnarray*}
\left\{ J_{\mu}\right\}  & = & \left\{ \rho,-\overrightarrow{J}\right\} .\end{eqnarray*}
The Lorentz four - force equation of motion for dyon is written as;

\begin{eqnarray}
f_{\mu}=m_{0}\ddot{x}_{\mu}= & Re & Q^{*}(G_{\mu\nu}u^{\nu})\label{eq:21}\end{eqnarray}
where $'Re'$ denotes the real part, $\{\ddot{x}_{\mu}\}$ is the
four - acceleration and $\left\{ u^{\nu}\right\} $ is the four -
velocity of the particle and $Q$ is the generalized charge of dyon.

\section{Magnetohydrodynamic (MHD) equations for dyons}

Three well known equations for magnetohydrodynamics (MHD) \cite{key-1}
are 

\begin{eqnarray}
\frac{\partial n}{\partial t}+\overrightarrow{\nabla} & .(n\,\overrightarrow{u})= & 0\label{eq:22}\end{eqnarray}

\begin{eqnarray}
\frac{\partial f}{\partial t} & +(\overrightarrow{v.}\,\overrightarrow{\nabla}f)+\frac{\vec{F}}{m}\frac{\partial f}{\partial v}= & \left(\frac{\partial f}{\partial t}\right)_{c}\label{eq:23}\end{eqnarray}

\begin{eqnarray}
mn\left[\frac{\partial\overrightarrow{u}}{\partial t}+(\overrightarrow{u}.\overrightarrow{\nabla})\,\overrightarrow{u}\right] & = & en\overrightarrow{F}+\triangle\overrightarrow{P}.\label{eq:24}\end{eqnarray}
Equation (\ref{eq:22}) is known as continuity equation, (\ref{eq:23})
is recalled as the Boltzmann equation while the equation (\ref{eq:24})
has been used for famous Ohm's law. In equations (\ref{eq:22} - \ref{eq:24}),
$n$ is the number density , $\overrightarrow{u}$ is the drift velocity
(fluid velocity), $\overrightarrow{v}$ is the particle velocity,
$f$ is the partition (distribution) function, $\overrightarrow{F}$
is the force acting on the particle, $m$ is the mass of the particle,
$c$ is used for collision, $e$ is the electric charge and $\triangle\overrightarrow{P}$
denotes the change in momentum due to collision. Let us discuss these
equations for the particles associated with the simultaneous existence
of electric and magnetic charges (dyons). To start with the familiar
continuity equation of hydrodynamics (\ref{eq:22}) with respect to
the conservation of matter (charge) we express\begin{eqnarray}
\frac{\partial n}{\partial t}+\overrightarrow{\nabla} & .(n\overrightarrow{u})= & 0\label{eq:25}\end{eqnarray}
For the case of electric charge, $n$ is replace by electric charge
density $\rho_{e}$. So the equation (\ref{eq:25}) takes the following
form 

\begin{eqnarray}
\frac{\partial\rho_{e}}{\partial t}+\overrightarrow{\nabla}.(\rho_{e}\overrightarrow{u}) & = & 0.\label{eq:26}\end{eqnarray}
Similarly on postulating the existence of magnetic charge (i.e. monopole),
we may replace $n$ by magnetic charge density $\rho_{m}$. Hence
the equation (\ref{eq:25}) reduces to the continuity equation in
presence of pure magnetic monopole i.e. \begin{eqnarray}
\frac{\partial\rho_{m}}{\partial t}+\overrightarrow{\nabla}.(\rho_{m}\overrightarrow{u}) & = & 0.\label{eq:27}\end{eqnarray}
As such, for the case of plasma for dyon (i.e; particle carrying both
electric and magnetic charge simultaneously), we may replace the number
density $n$ by generalized source density $\rho=\rho_{e}-i\rho_{g}$
of dyons. So using equations (\ref{eq:25} - \ref{eq:27}), we get

\begin{eqnarray}
\frac{\partial\rho}{\partial t}+\overrightarrow{\nabla}.(\rho\overrightarrow{u}) & = & 0\label{eq:28}\end{eqnarray}
which represents the generalized form of continuity equation (first
magnetohydrodynamics (MHD) equation) for dyonic plasma. 

Similarly, we may express the second equation of magnetohydrodynamics
(MHD) for the plasma of dyons. For this we replace the force $\overrightarrow{F}$
of equation (\ref{eq:23}) as the force exerting on the particle simultaneously
carrying electric and magnetic charges (dyons). The force of dyons
given by equation (\ref{eq:21}) reduces to 

\begin{eqnarray}
\overrightarrow{F}=e(\overrightarrow{E}+\overrightarrow{v}\times\overrightarrow{H})+g(\overrightarrow{H} & -\overrightarrow{v}\times\overrightarrow{E} & ).\label{eq:29}\end{eqnarray}
Substituting the equation (\ref{eq:29}) for force $\overrightarrow{F}$
of dyons into equation (\ref{eq:23}) and multiplying by $m\overrightarrow{v}$
and then integrating it over the velocity $d\overrightarrow{v}$,
we get,

\begin{align}
m\int\overrightarrow{v}\frac{\partial f}{\partial t}d\overrightarrow{v} & +m\int\overrightarrow{v}(\overrightarrow{v}.\overrightarrow{\nabla}f)d\overrightarrow{v}\nonumber \\
+ & \int\overrightarrow{v}[e(\overrightarrow{E}+\overrightarrow{v}\times\overrightarrow{H})+g(\overrightarrow{H}-\overrightarrow{v}\times\overrightarrow{E})]\frac{\partial f}{\partial v}d\overrightarrow{v}=\int m\overrightarrow{v}\left(\frac{\partial f}{\partial t}\right)_{c}d\overrightarrow{v}.\label{eq:30}\end{align}
The first term of equation (\ref{eq:30}) reduces to

\begin{eqnarray}
m\int\overrightarrow{v}\frac{\partial f}{\partial t}d\overrightarrow{v}=m\frac{\partial}{\partial t}\int\overrightarrow{v}fd\overrightarrow{v}=m\frac{\partial}{\partial t} & (n\overrightarrow{u}) & .\label{eq:31}\end{eqnarray}
Similarly, the second term of equation (\ref{eq:30}) is expressed
as,

\begin{equation}
m\int\overrightarrow{v}(\overrightarrow{v}.\overrightarrow{\nabla}f)d\overrightarrow{v}=m\int(\overrightarrow{\nabla}.f\overrightarrow{v})\,\overrightarrow{v}d\overrightarrow{v}=m\int(\overrightarrow{\nabla}.f\overrightarrow{v\,})\overrightarrow{v}d\overrightarrow{v}\label{eq:32}\end{equation}
where 

\begin{eqnarray}
\int(\overrightarrow{\nabla}.f\overrightarrow{v}) & \overrightarrow{v}d\overrightarrow{v}=(\overrightarrow{\nabla}.\overline{n\overrightarrow{v}})\overrightarrow{v} & .\label{eq:33}\end{eqnarray}
Expressing the particle velocity $\overrightarrow{v}$ in terms of
the average velocity (fluid velocity) $\overrightarrow{u}$ and thermal
velocity $\overrightarrow{w}$, as 

\begin{eqnarray}
\overrightarrow{v} & =\overrightarrow{w}+ & \overrightarrow{u}\label{eq:34}\end{eqnarray}
 we get \begin{eqnarray}
(\overrightarrow{\nabla}.\overline{n\overrightarrow{v}})\overrightarrow{v}=(\overrightarrow{\nabla}.\overline{n\overrightarrow{u}})\overrightarrow{u}+(\overrightarrow{\nabla}.\overline{n\overrightarrow{w}})\overrightarrow{w}+ & 2 & (\overrightarrow{\nabla}.\overline{n\overrightarrow{u}})\overrightarrow{w}\label{eq:35}\end{eqnarray}
where $\overline{w}$ the average thermal velocity vanishes. The quantity
$m(\overline{n\overrightarrow{w}})\overrightarrow{w}$ is expressed
as the stress tensor $I$. Then the first term of equation (\ref{eq:35})
reduces to\begin{eqnarray}
(\overrightarrow{\nabla}.\overline{n\overrightarrow{u}})\overrightarrow{u} & =\overrightarrow{u}(\overrightarrow{\nabla}.n\overrightarrow{u})+ & n(\overrightarrow{u}.\overrightarrow{\nabla})\overrightarrow{u}.\label{eq:36}\end{eqnarray}
As such, equation (\ref{eq:32}) reduces to

\begin{eqnarray}
m\int\overrightarrow{v}(\overrightarrow{v}.\overrightarrow{\nabla}f)d\overrightarrow{v}=m\overrightarrow{u}\overrightarrow{\nabla}.(nu)+ & mn(\overrightarrow{u}.\overrightarrow{\nabla})\overrightarrow{u}+ & \overrightarrow{\nabla}.I.\label{eq:37}\end{eqnarray}
So, the third term of equation (\ref{eq:30}) is expressed as 

\begin{equation}
e\int\overrightarrow{v}(\overrightarrow{E}+\overrightarrow{v}\times\overrightarrow{H})\frac{\partial f}{\partial v}d\overrightarrow{v}=-en_{e}(\overrightarrow{E}+\overrightarrow{v}\times\overrightarrow{H}).\label{eq:38}\end{equation}
Similarly the fourth term of equation (\ref{eq:30}) reduces to

\begin{equation}
g\int\overrightarrow{v}(\overrightarrow{H}-\overrightarrow{v}\times\overrightarrow{E})\frac{\partial f}{\partial t}d\overrightarrow{v}=-gn_{m}(\overrightarrow{H}-\overrightarrow{v}\times\overrightarrow{E}).\label{eq:39}\end{equation}
On the other hand the right hand side term of equation (\ref{eq:30})
is analogous to the change in momentum $\triangle\overrightarrow{P}$
due to collision i.e;

\begin{equation}
\triangle\overrightarrow{P}=\int mv\left(\frac{\partial f}{\partial t}\right)_{c}d\overrightarrow{v}.\label{eq:40}\end{equation}
Using equations (\ref{eq:31}, \ref{eq:37}, \ref{eq:38}, \ref{eq:39})
and (\ref{eq:40}) , we get the following reduced form of equation
(\ref{eq:30}) i.e.

\begin{eqnarray}
mn\left[(\frac{\partial\overrightarrow{u}}{\partial t})+(\overrightarrow{u}.\overrightarrow{\nabla})\overrightarrow{u}\right]=\rho^{\star}\overrightarrow{\psi}+\overrightarrow{J}\times\overrightarrow{\psi}+ & \triangle\overrightarrow{P}+ & \overrightarrow{\nabla}.I,\label{eq:41}\end{eqnarray}
which is the Boltzmann equation the case of dyonic plasma. In equation
(\ref{eq:41}) $\rho^{\star}$ is the complex conjugate of the dyonic
charge source density, $\vec{\psi}$ is complex vector field of dyon,
$\overrightarrow{J}$ is current source density of dyon, $n_{e}$
is the number density of electric charge, $n_{m}$ is the number density
of magnetic charge. Physical interpretation of equation (\ref{eq:41})
has thus been described due to the scattering between either purely
electric or magnetic charge carriers. The equation (\ref{eq:41})
is thus described as the fluid equation of motion for generalized
field of dyons (i.e. dyonic fluid) which is the modified form of second
equation for magnetohydrodynamics (MHD) in case of dyonic plasma. 

Third equation of magnetohydrodynamics (MHD) given by equation (\ref{eq:24})
is expressed as the generalized Ohm's law. We may write magnetohydrodynamics
(MHD) equation (\ref{eq:24}) in presence of electric charge as

\begin{eqnarray}
mn_{e}\left[\frac{\partial u_{e}}{\partial t}+u_{e}\nabla u_{e}\right] & = & en_{e}[\vec{E}+\vec{v}\times\vec{H}]+\triangle\overrightarrow{P}_{e}.\label{eq:42}\end{eqnarray}
Accordingly, the magnetohydrodynamics (MHD) equation (\ref{eq:24})
in presence of pure magnetic charge may be expressed as

\begin{eqnarray}
mn_{g}\left[\frac{\partial u_{g}}{\partial t}+u_{g}\nabla u_{g}\right] & = & gn_{g}[n_{m}\vec{H}-\vec{v}\times\vec{E}]+\triangle\overrightarrow{P}_{g}.\label{eq:43}\end{eqnarray}
Multiplying equation (\ref{eq:42}) by $\frac{e}{m}$ and equation
(\ref{eq:43}) by $\frac{ig}{m}$ and then subtracting , we get

\begin{equation}
en_{e}\frac{\partial u_{e}}{\partial t}-ign_{g}\frac{\partial u_{g}}{\partial t}=\frac{e^{2}n_{e}}{m}[\vec{E}+\vec{v}\times\vec{H}]+\frac{e}{m}\triangle\overrightarrow{P}_{e}-i\frac{g^{2}n_{g}}{m}[\vec{H}-\vec{v}\times\vec{E}]-i\frac{g}{m}\triangle\overrightarrow{P}_{g},\label{eq:44}\end{equation}
where 

\begin{eqnarray}
j_{e} & = & en_{e}u_{e};\nonumber \\
j_{m} & = & gn_{g}u_{g}.\label{eq:45}\end{eqnarray}
Using equations (\ref{eq:44} - \ref{eq:45}), we get

\begin{equation}
\frac{\partial J}{\partial t}=\frac{1}{m}[e^{2}n_{e}F_{e}-ig^{2}n_{g}F_{m}]+\frac{1}{m}(e\triangle\overrightarrow{P_{e}}-ig\triangle\overrightarrow{P_{m}})\label{eq:46}\end{equation}
where $J=j_{e}-ij_{g}$ is used for the generalized current density
for dyons, $F_{e}=e(\overrightarrow{E}+\overrightarrow{u}\times\overrightarrow{H})$
is the force acting on the electric charged particle and $F_{m}=g(\overrightarrow{H}-\overrightarrow{u}\times\overrightarrow{E})$
describes the force acting on the magnetic charged particle. As such,
the equation (\ref{eq:46}) is the modified form of the third equation
of magnetohydrodynamics (MHD) for the case of generalized fields of
dyons (i.e. dyonic plasma). In the absence of the $\overrightarrow{H}=0$
($\overrightarrow{E}=0)$ all the three modified equations (\ref{eq:28},
\ref{eq:41} \& \ref{eq:46}) of magnetohydrodynamics (MHD) associated
with dyonic plasma reduces to the usual differential equation of magnetohydrodynamics
(MHD) in presence of electric ( magnetic) charge only. The equation
(\ref{eq:46}) provides the combination of two Ohm's law. It is due
to the mixed plasma of dyons as the consequence of presence of electric
and magnetic charges on dyons. In this case we have considered low
wavelength approximation. Consequently this case the electron - ion
and magnetic monopole - magneto ions recombine due to short distance
effects. If the plasma dynamics becomes too fast, resonances occur
with the motions of individual particles which invalidate the MHD
equations. Furthermore, effects, such as particle inertia and the
Hall effect, which are not taken into account in the MHD equations,
become important. Since MHD is a single fluid plasma theory, a single
fluid approach is justified because the perpendicular motion is dominated
by $\overrightarrow{E}\times\overrightarrow{H}$ drifts. For the case
of slow plasma dynamics, the motions of the dyon and ion fluids become
sufficiently different as single fluid approach is no longer tenable.
This also occurs whenever the diamagnetic velocities, which are quite
different for different plasma species, become comparable to the $\overrightarrow{E}\times\overrightarrow{H}$
velocity . Furthermore, effects such as plasma resistivity, viscosity,
and thermal conductivity, which are not taken into account in the
MHD equations, become important in this case.

\section{Frequency of Dyonic Plasma}

The case of dyonic plasma is not the case of single particle motions
but rather collective motion of the various charge species of dyonic
plasma (i.e. electric plasma and monopole plasma). So, the first,
and most important is to discuss the electrostatic plasma oscillation
responsible for plasma frequency. These oscillations occur when one
of the species is displaced from the other. 

For this we start with the equation of motion for a particle which
has electric charge $e$ and mass $m$ in an electric field as

\begin{eqnarray}
m\overrightarrow{\ddot{r}} & = & e\overrightarrow{E}.\label{eq:47}\end{eqnarray}
where $\overrightarrow{\ddot{r}}$ denotes the acceleration. If $\overrightarrow{E}=\overrightarrow{E}\, e^{i(\overrightarrow{k}.\overrightarrow{r}-\omega t)}$
is expressed as an incident plane wave, then equation (\ref{eq:47})
reduces to

\begin{eqnarray}
m\overrightarrow{\dot{r}} & = & -\frac{e\overrightarrow{E}}{i\omega}.\label{eq:48}\end{eqnarray}
So the electric current source density is expressed as 

\begin{eqnarray}
\overrightarrow{j_{e}} & = & n_{e}e\overrightarrow{\dot{r}}=-\frac{n_{e}e^{2}\overrightarrow{E}}{im\omega}=\delta_{e}\overrightarrow{E}\label{eq:49}\end{eqnarray}
where

\begin{eqnarray}
\delta_{e} & = & -\frac{n_{e}e^{2}}{im\omega}.\label{eq:50}\end{eqnarray}
Similarly for a particle with monopole of charge $g$ moving in a
magnetic field, we have

\begin{eqnarray}
\overrightarrow{j_{m}} & = & \delta_{m}\overrightarrow{H}\label{eq:51}\end{eqnarray}
where 

\begin{eqnarray}
\delta_{m} & = & -\frac{n_{g}g^{2}}{im\omega}.\label{eq:52}\end{eqnarray}
With the help of Maxwell's Dirac equation (\ref{eq:1}) and using
the relations (\ref{eq:45}), (\ref{eq:49}- \ref{eq:52}), we get\begin{eqnarray}
k^{2} & = & \omega^{2}-\frac{n_{e}}{m}qq*+\left(\frac{n_{e}eg}{m\omega}\right)^{2}=\omega^{2}-\omega_{p}^{2}\label{eq:53}\end{eqnarray}
where

\begin{eqnarray}
\omega_{p}^{2} & = & \left(\frac{n_{e}}{m}qq*-\left(\frac{n_{e}eg}{m\omega}\right)^{2}\right)\label{eq:54}\end{eqnarray}
where $q=e-ig$ is the generalized charge of dyon. In equation (\ref{eq:54})
$\omega_{p}$ is the plasma frequency of the dyon, $q^{\star}$ is
the complex conjugate of generalized charge $q$ of dyon and $k$
is known as the usual wave number. There is a different plasma frequency
for each species. For $k$ to be real, only those generalized electromagnetic
waves are allowed to pass, for which $\omega>\omega_{p}$. At very
high frequencies, $\omega=ck$, dyon can not respond fast enough,
and plasma effects are negligible. Thus Plasma frequency sets the
lower cuts for the frequencies of electromagnetic radiation that can
pass through a plasma. The metals shine by reflecting most of light
in visible range. The visible light can not pass through the metal
because the plasma frequency of electrons in metal falls in ultraviolet
region. For frequencies in ultra voilet region ($UV$), metals are
transparent. The earth's ionosphere reflects radio waves in the same
reason. The electron (monopole) densities at various heights in the
ionosphere can be inered by studying the reflection of pulses of radiation
transmitted vertically upwards. Also, the broadcast of various radio
signals in communication on earth is possible only because of reflection
from the ionosphere.

\section{Magnetohydrodynamic (MHD) waves for dyons}

Plasma is a complex fluid that support many plasma wave modes. Restoring
forces include kinetic pressure and electromagnetic forces \cite{key-12}.
Let us investigate the small amplitude waves propagating through a
spatially uniform MHD plasma. For this, we take the two cases. Case
I- we define .

\begin{eqnarray}
\overrightarrow{E}+\overrightarrow{v}\times\overrightarrow{H} & = & 0\label{eq:55}\end{eqnarray}
which describes the Ohm's law for the dynamics of electric charge.
Similarly for case -II we have 

\begin{eqnarray}
\overrightarrow{H}-\overrightarrow{v}\times\overrightarrow{E} & = & 0\label{eq:56}\end{eqnarray}
which may be identified as the Ohm's law for free magnetic monopole.
So, the force equation is expressed as 

\begin{eqnarray}
\rho\frac{\partial\overrightarrow{v}}{\partial t}+\rho(\overrightarrow{v}.\overrightarrow{\nabla})\overrightarrow{v} & = & (\overrightarrow{\nabla}\times\overrightarrow{H})\times\overrightarrow{H}+(\overrightarrow{\nabla}\times\overrightarrow{E})\times\overrightarrow{E}+\overrightarrow{\nabla}p\label{eq:57}\end{eqnarray}
while the equation of state is given as

\begin{eqnarray}
\overrightarrow{\nabla}p & = & -V_{s}^{2}\overrightarrow{\nabla}\rho\label{eq:58}\end{eqnarray}
where $V_{s}$ is the speed of the dyon. Substituting equation (\ref{eq:58})
into equation (\ref{eq:57}) we may write the force equation as

\begin{eqnarray}
\rho\frac{\partial\overrightarrow{v}}{\partial t}+\rho(\overrightarrow{v}.\overrightarrow{\nabla})\overrightarrow{v} & = & (\overrightarrow{\nabla}\times\overrightarrow{H})\times\overrightarrow{H}+(\overrightarrow{\nabla}\times\overrightarrow{E})\times\overrightarrow{E}-V_{s}^{2}\overrightarrow{\nabla}\rho.\label{eq:59}\end{eqnarray}
Post multiplying vectorially the fourth Maxwell's Dirac equation (\ref{eq:1})
by $\overrightarrow{H}$ and rearranging the terms, we get,

\begin{eqnarray}
\overrightarrow{j_{e}}\times\overrightarrow{H} & = & (\overrightarrow{\nabla}\times\overrightarrow{H})\times\overrightarrow{H}-\frac{\partial\overrightarrow{E}}{\partial t}\times\overrightarrow{H}.\label{eq:60}\end{eqnarray}
Similarly post multiplying vectorially the third Maxwell's Dirac equation
(\ref{eq:1}) by $\overrightarrow{E}$, we get

\begin{eqnarray}
\overrightarrow{j_{m}}\times\overrightarrow{E} & =- & (\overrightarrow{\nabla}\times\overrightarrow{E})\times\overrightarrow{E}-\frac{\partial\overrightarrow{H}}{\partial t}\times\overrightarrow{E}.\label{eq:61}\end{eqnarray}
Hence the equation (\ref{eq:41}) reduces to 

\begin{align}
mn\left[(\frac{\partial\overrightarrow{v}}{\partial t})+(\overrightarrow{v}.\overrightarrow{\nabla})\overrightarrow{v}\right]= & en_{e}\overrightarrow{E}+gn_{g}\overrightarrow{H}\nonumber \\
+ & (\overrightarrow{\nabla}\times\overrightarrow{H})\times\overrightarrow{H}+(\overrightarrow{\nabla}\times\overrightarrow{E})\times\overrightarrow{E}+\triangle P+\frac{\partial}{\partial t}(\overrightarrow{H}\times\overrightarrow{E})+\overrightarrow{\nabla}.\overrightarrow{I}.\label{eq:62}\end{align}
With the help of equations (\ref{eq:55}), (\ref{eq:56}) and equation
(\ref{eq:1}), we get

\begin{eqnarray}
\frac{\partial\overrightarrow{H}}{\partial t} & = & \overrightarrow{\nabla}\times(\overrightarrow{v}\times\overrightarrow{H})-\overrightarrow{j_{m}};\label{eq:63}\\
\frac{\partial\overrightarrow{E}}{\partial t} & = & \overrightarrow{\nabla}\times(\overrightarrow{v}\times\overrightarrow{E})-\overrightarrow{j_{e}}.\label{eq:64}\end{eqnarray}
Applying the perturbation transformations about the equilibrium values
i.e. 

\begin{align}
\rho & \longmapsto\rho_{0}+\rho_{1};\nonumber \\
H & \longmapsto H_{0}+H_{1};\nonumber \\
E & \longmapsto E_{0}+E_{1};\nonumber \\
v & \longmapsto v_{1};\label{eq:65}\end{align}
where $ρ_{0}$ is the background density of the unperturbed fluid
and $v_{0}=0$ (i.e. the fluid is at rest). As such, with the help
of equation (\ref{eq:65}), we get the following reduced expressions
for equations (\ref{eq:28}) and (\ref{eq:62}) 

\begin{align}
\frac{\partial\rho_{1}}{\partial t}+\rho_{0}\overrightarrow{\nabla}.\overrightarrow{v_{1}} & =0;\nonumber \\
\frac{\partial^{2}\overrightarrow{v_{1}}}{\partial t^{2}}+V_{s}^{2}\overrightarrow{\nabla}\left[-\overrightarrow{\nabla}.\overrightarrow{v_{1}}\right] & +\frac{H_{0}}{\rho_{0}}\left[\overrightarrow{\nabla}\times\left\{ \overrightarrow{\nabla}\times\left(\overrightarrow{v_{1}}\times\overrightarrow{H_{0}}\right)-\overrightarrow{j_{m}}\right\} \right]\nonumber \\
+ & \frac{E_{0}}{\rho_{0}}\left[\overrightarrow{\nabla}\times\left\{ \overrightarrow{\nabla}\times\left(\overrightarrow{v_{1}}\times\overrightarrow{E_{0}}\right)-\overrightarrow{j_{e}}\right\} \right]=0.\label{eq:66}\end{align}
Here we have used

\begin{eqnarray}
V_{A} & = & \frac{H_{0}}{\left(\rho_{0}\right)^{\frac{1}{2}}};\,\,\,\,\,\, V_{B}=\frac{E_{0}}{\left(\rho_{0}\right)^{\frac{1}{2}}};\label{eq:67}\end{eqnarray}

\begin{eqnarray}
v_{1}(\overrightarrow{r},t) & = & v_{0}\exp i(\overrightarrow{k}.\overrightarrow{r}-\omega t);\label{eq:68}\end{eqnarray}
\begin{eqnarray}
\overrightarrow{\nabla}\longrightarrow i\overrightarrow{k} & and & \frac{\partial}{\partial t}\longrightarrow-i\omega.\label{eq:69}\end{eqnarray}
Hence the dispersion relation (\ref{eq:66}) takes the following form 

\begin{align}
-\omega^{2}v_{1}+V_{s}^{2}(\overrightarrow{k}.\overrightarrow{v_{1}})k-\overrightarrow{V_{A}}\times & \left[\overrightarrow{k}\times[\overrightarrow{k}\times(\overrightarrow{v_{1}}\times\overrightarrow{V_{A}})+\frac{\overrightarrow{j_{m}}}{(\rho_{0})^{\frac{1}{2}}}\right]\nonumber \\
- & \overrightarrow{V_{B}}\times\left[\overrightarrow{k}\times[\overrightarrow{k}\times(\overrightarrow{v_{1}}\times\overrightarrow{V_{B}})+\frac{\overrightarrow{j_{e}}}{(\rho_{0})^{\frac{1}{2}}}\right]=0.\label{eq:70}\end{align}
Expanding the equation (\ref{eq:70}) and using vector triple product,
we get

\begin{align}
-\omega^{2}v_{1}+ & [V_{s}^{2}+V_{A}^{2}+V_{B}^{2}](\overrightarrow{k}.\overrightarrow{v_{1}})k\nonumber \\
+ & (\overrightarrow{k}.\overrightarrow{V_{A}})[(\overrightarrow{k}.\overrightarrow{V_{A}})\overrightarrow{v_{1}}-(\overrightarrow{V_{A}}.\overrightarrow{v_{1}})\overrightarrow{k}-(\overrightarrow{k}.\overrightarrow{v_{1}})\overrightarrow{V_{A}}]-\frac{\overrightarrow{V_{A}}\times(\overrightarrow{k}\times\overrightarrow{j_{m}})}{(\rho_{0})^{\frac{1}{2}}}\nonumber \\
+ & (\overrightarrow{k}.\overrightarrow{V_{B}})[(\overrightarrow{k}.\overrightarrow{V_{B}})\overrightarrow{v_{1}}-(\overrightarrow{V_{B}}.\overrightarrow{v_{1}})\overrightarrow{k}-(\overrightarrow{k}.\overrightarrow{v_{1}})\overrightarrow{V_{B}}]-\frac{\overrightarrow{V_{B}}\times(\overrightarrow{k}\times\overrightarrow{j_{e}})}{(\rho_{0})^{\frac{1}{2}}}=0.\label{eq:71}\end{align}
It describes the case of a kind of dyonoacoustic wave. Substituting
$\overrightarrow{k}\perp\overrightarrow{H_{0}}$ , $\overrightarrow{k}\perp\overrightarrow{E_{0}}$
and $\overrightarrow{j_{m}}\parallel\overrightarrow{V_{A}},$\ $\overrightarrow{j_{e}}\parallel\overrightarrow{V_{B}}$
, we get 

\begin{eqnarray}
\overrightarrow{k}.\overrightarrow{V_{A}} & = & \overrightarrow{k}.\overrightarrow{V_{B}}=0\label{eq:72}\end{eqnarray}
and 

\begin{eqnarray}
\overrightarrow{V_{A}}\times\overrightarrow{j_{m}}=\overrightarrow{V_{B}}\times\overrightarrow{j_{e}} & = & 0.\label{eq:73}\end{eqnarray}
As such, we get the following expression for the dispersion relation 

\begin{eqnarray}
-\omega^{2}v_{1}+[V_{s}^{2}+V_{A}^{2}+V_{B}^{2}](\overrightarrow{k}.\overrightarrow{v_{1}})k & = & 0.\label{eq:74}\end{eqnarray}
The vector nature of the equation (\ref{eq:74}) requires that the
perturbed fluid velocity $\overrightarrow{v_{1}}$ must be parallel
to the propagation direction $\overrightarrow{k}$ so that $\overrightarrow{k}.\overrightarrow{v_{1}}=kv_{1}$.
Thus the wave is longitudinal in nature and its dispersion relation
becomes

\begin{eqnarray}
v_{\phi} & =\frac{\omega}{k} & =V_{s}^{2}+V_{A}^{2}+V_{B}^{2}.\label{eq:75}\end{eqnarray}
Therefore the dyon acoustic waves propagating with velocity $v_{\phi}$
in this case. This is known as the dyonoacoustic, dyonosonic or simply
compressional wave which involves compression and rarefaction for
the electromagnetic lines of force along with plasma oscillations.

\section{Energy of dyons}

The energy of the dyonic plasma is related with the dispersive properties
of the wave oscillations. Starting from first principle for the electromagnetic
energy density and taking into account the specific features of dispersive
relations of electromagnetic waves, we may obtain the expression for
electromagnetic energy density (namely the Poynting Theorem) \cite{key-12}.
From the third and fourth Maxwell's Dirac equation (\ref{eq:1}),
we obtain 

\begin{eqnarray}
\overrightarrow{E}.(\overrightarrow{\nabla}\times\overrightarrow{H})-\overrightarrow{H}.(\overrightarrow{\nabla}\times\overrightarrow{E}) & = & \overrightarrow{E}.\frac{\partial\overrightarrow{E}}{\partial t}+\overrightarrow{H}.\frac{\partial\overrightarrow{H}}{\partial t}+\overrightarrow{j_{e}}.\overrightarrow{E}+\overrightarrow{j_{m}}.\overrightarrow{H}.\label{eq:76}\end{eqnarray}
This equation may then be written as the conservation law of energy
as 

\begin{eqnarray}
\frac{\partial W}{\partial t}+\overrightarrow{\nabla}.\overrightarrow{P} & = & 0\label{eq:77}\end{eqnarray}
where 

\begin{eqnarray}
\overrightarrow{P} & = & \overrightarrow{E}\times\overrightarrow{H}\label{eq:78}\end{eqnarray}
is called the Poynting vector. The rate of change of the energy density
$\frac{\partial W}{\partial t}$ is then defined as 

\begin{eqnarray}
\frac{\partial W}{\partial t} & = & \overrightarrow{E}.\frac{\partial\overrightarrow{E}}{\partial t}+\overrightarrow{H}.\frac{\partial\overrightarrow{H}}{\partial t}+\overrightarrow{j_{e}}.\overrightarrow{E}+\overrightarrow{j_{m}}.\overrightarrow{H}\label{eq:79}\end{eqnarray}
so that we may obtain the energy density by taking time integration
as

\begin{align}
W(t)= & W_{0}(t)+\int_{t_{0}}^{t}dt[\overrightarrow{E}.\frac{\partial\overrightarrow{E}}{\partial t}+\overrightarrow{H}.\frac{\partial\overrightarrow{H}}{\partial t}+\overrightarrow{j_{e}}.\overrightarrow{E}+\overrightarrow{j_{m}}.\overrightarrow{H}]\nonumber \\
= & W_{0}(t)+\left[\frac{E_{0}^{2}+H_{0}^{2}}{2}\right]+\int_{t_{0}}^{t}dt(\overrightarrow{j_{m}}.\overrightarrow{H}+\overrightarrow{j_{e}}.\overrightarrow{E})\label{eq:80}\end{align}
where $W_{0}(t)$ is the energy density at reference point $t_{0}$.
The quantity $(\overrightarrow{j_{m}}.\overrightarrow{H}+\overrightarrow{j_{e}}.\overrightarrow{E})$
is the rate of change of kinetic energy density of the dyon. This
can be seen by taking the dot product of force equation which is taking
this form with $\overrightarrow{v}$ 

\begin{equation}
m\overrightarrow{v}.\frac{d\overrightarrow{v}}{dt}=\overrightarrow{v}.(\overrightarrow{F_{e}}+\overrightarrow{F_{m}})=\overrightarrow{v}.[e(\overrightarrow{E}+\overrightarrow{v}\times\overrightarrow{H})+g(\overrightarrow{H}-\overrightarrow{v}\times\overrightarrow{E})]\label{eq:81}\end{equation}
which may also be written as 

\begin{eqnarray}
\frac{d}{dt}(\frac{1}{2}mv^{2}) & = & e\overrightarrow{E}.\overrightarrow{v}+g\overrightarrow{H}.\overrightarrow{v}.\label{eq:82}\end{eqnarray}
Since this is the rate of change of kinetic energy of a single dyon,
the rate of change of the kinetic energy density $T$ for the entire
system of dyons is found by summing over the energies of the individual
dyons i.e.

\begin{eqnarray}
\frac{d}{dt}(T) & = & \sum_{i}\int dvf_{i}(e_{i}\overrightarrow{E}.\overrightarrow{v}+g_{i}\overrightarrow{H}.\overrightarrow{v})=\overrightarrow{E}.\overrightarrow{j_{e}}+\overrightarrow{H}.\overrightarrow{j_{m}}=Re(\overrightarrow{J}.\overrightarrow{\psi^{\star}}).\label{eq:83}\end{eqnarray}
This shows that positive value of $(\overrightarrow{E}.\overrightarrow{j_{e}}+\overrightarrow{H}.\overrightarrow{j_{m}})$
corresponds to the increases the kinetic energy of dyons whereas the
negative value of $(\overrightarrow{E}.\overrightarrow{j_{e}}+\overrightarrow{H}.\overrightarrow{j_{m}})$
corresponds to the decrease of kinetic energy of dyons. The latter
situation is possible only if the dyon starts working with a finite
initial kinetic energy density. Since $(\overrightarrow{E}.\overrightarrow{j_{e}}+\overrightarrow{H}.\overrightarrow{j_{m}})$
accounts for the changes in the dyon kinetic energy density, $W$
must be the sum of the generalized electromagnetic field density and
the particle energy density. In our case, we have considered a fluid
element of dyonic plasma for which the overall charge is taken to
be neutral. So, an external electromagnetic field cannot cause motion
of a fluid element as a whole, but will sets up currents due to the
motion of opposite charges in opposite directions. Due to these currents,
an external electromagnetic field exerts a force on the fluid element
and changes its direction of motion. Here we have described the motion
of plasma oscillations for two different fluids associated with the
electric and magnetic charges or a composite system of dyons.

\textbf{\underbar{ACKNOWLEDGMENT}}\textbf{: }One of PSB is thankful
to Chinese Academy of Sciences and Third World Academy of Sciences
for awarding him CAS - TWAS Visiting Scholar Fellowship to pursue
a research program in China. He is also grateful to Professor Tianjun
Li for his hospitality at Institute of Theoretical Physics, Beijing,
China.

\end{document}